\documentclass[a4paper,11pt]{article}
\pdfoutput=1 

\usepackage{jheppub} 
\usepackage[english]{babel}

\usepackage[nottoc]{tocbibind}

\usepackage[T1]{fontenc} 
\usepackage{amsmath}
\usepackage{amsfonts}
\usepackage{amssymb}
\usepackage{hepunits}
\usepackage[font=small,skip=0pt]{caption}
\usepackage{float}
\usepackage[utf8]{inputenc}
\usepackage[colorinlistoftodos]{todonotes}
\usepackage[nottoc]{tocbibind}

\title{\boldmath Binary Collisions of Dark Matter Blobs}

\author{Melissa~ D.~ Diamond*,\note[*]{corresponding author} David~ E.~ Kaplan,  Surjeet Rajendran}

\affiliation{Department of Physics $\&$ Astronomy,\\The Johns Hopkins University, Baltimore, MD 21218, USA\\
Department of Physics\\}

\emailAdd{mdiamon8@jhu.edu, david.kaplan@jhu.edu,  sranjend4@jhu.edu}

\begin{document}



\vspace*{1cm}

\abstract{We describe the model-independent mechanism by which dark matter and dark matter structures heavier than $\sim 8\times 10^{11}$ GeV form binary pairs in the early Universe that spin down and merge both in the present and throughout the Universe's history, producing potentially  observable signals.  Sufficiently dense dark objects will dominantly collide through binary mergers instead of random collisions. We detail how one would estimate the merger rate accounting for finite size effects, multibody interactions, and friction with the thermal bath. We predict how mergers of dark dense objects could be detected through gravitational and electromagnetic signals, noting that such mergers could be a unique source of high frequency gravitational waves. We rule out objects whose presence would contradict observations of the CMB and diffuse gamma-rays. }
\maketitle

\section{Introduction}
Understanding  dark matter has been a central goal of particle physics and cosmology for decades.  Extensive experimental and observational efforts have been made to this end.  Particle dark matter has been searched for in particle colliders \cite{ATLAS:2019wdu,Vannerom:2019p2,Buchmueller:2017qhf,Bai:2010hh}, direct detection experiments \cite{SuperCDMS:2017nns,CRESST:2019jnq,CoGeNT:2012sne,EDELWEISS:2017lvq,XENON:2017lvq,Adhikari:2018ljm,Behnke:2016lsk,DRIFT:2016utn,SENSEI:2020dpa}, and through indirect searches  \cite{Conrad:2015bsa,Zitzer:2017xlo,IceCube:2021kuw,Habig:2001ei}.  Ultra-light dark matter, such as axions and axion like particles, can be probed through its interaction with astrophysical objects \cite{Foster:2020pgt} and the cosmic microwave background (CMB) \cite{Fedderke:2019ajk} and through direct detection experiments \cite{ADMX:2019uok,HAYSTAC:2018rwy,DellaValle:2014wea,Ehret:2010mh,CAST:2017uph,McAllister:2017lkb,XENON100:2014csq}.  On the opposite end of the mass spectrum are primordial black holes (PBHs) and massive compact halo objects (MACHOs) with masses above $10^{21}$g.  These have been thoroughly explored and generally ruled out by friction effects \cite{10.1093/mnras/259.1.27P}, clustering \cite{Carr:2018rid,Brandt:2016aco}, and  lensing searches \cite{Barnacka_2012,Niikura:2017zjd,Niikura:2019kqi,EROS-2:2006ryy,Oguri:2017ock}, though less extensively for MACHOs. 

Despite the considerable programs dedicated to understanding dark matter, little has been done to probe models with masses between the Planck mass ($m_p\sim2\times 10^{-5}$ g) and the lower mass limit for lensing searches ($\sim10^{21}$g).  This is not for lack of reasonable dark matter models in this mass range. Many have been proposed, including nuclear dark matter  \cite{Brandt:2016aco, Hardy:2015boa,Wise:2014jva, Gresham:2017zqi,Bai:2018dxf}, stabilized black holes and their remnants \cite{Lehmann:2019zgt,Diamond:2021scl}, and exotic compact objects (ECOs) such as boson stars and fermion stars \cite{Giudice:2016zpa}, among others \cite{Lennon:2017tqq,Berezinsky:2013fxa}.  This region of parameter space remains largely unexplored because it is difficult to probe.  Present and planned particle colliders on Earth cannot produce particles heavier than $\sim$ 100 TeV \cite{FCC:2018bvk}.  Direct detection efforts are limited by low number densities, while lensing searches are thwarted by low masses \cite{Carr:2020gox}.  Dark matter searches in this mass window are limited to indirect probes of its self-interactions or its interactions with astrophysical objects.

Fortunately, much of the dark matter in this difficult to probe mass range, and many masses outside of it, has a simple, reasonably model-independent self-interaction that can produce an observable signal.  Cold blobs of dark matter heavier than $8\times 10^{11}$ GeV and dense enough to behave as point-like objects form binary pairs in the early Universe.  Blobs with small initial separations can decouple from the Hubble flow and begin falling toward each other.   In-falling pairs of blobs are perturbed by tidal interactions with the next nearest blob, averting a head on collision and forming a binary.  These binaries spin down and merge, with many collisions taking place in the Milky Way today and in the observable past.  This phenomenon has been explored extensively for PBHs  \cite{Eroshenko:2016hmn,Sasaki:2016jop,Ali-Haimoud:2017rtz,Sasaki:2018dmp,Raidal:2018bbj,Jedamzik:2020ypm}, though never applied to other compact objects.  The mergers may produce observable gravitational or electromagnetic (EM) signals, some of which we describe below.    The details of this signal often depend on the blob's coupling to the Standard Model (SM).  We do not explore such model-dependent features of blobs, including how they form.  Regardless, the basic dynamics of binary mergers are shared for many dark matter blobs over a vast range of masses.  

In this paper, we describe a generic behavior of blob dark matter, the formation of binaries that merge at a high rate today, and how this can be used to predict and search for distinct exotic signals.  The paper is laid out as follows. Section \ref{sec:Rates} describes the general method for estimating the merger rate, as well as how to account for blobs of finite size,  with couplings to the SM, and with charged self-interactions.  Section \ref{sec:Signal} describes some general gravitational and EM signals that may result from these mergers and rules out blobs that produce enough EM radiation to alter current observations of the CMB anisotropy angular power spectrum and gamma-rays.  Section \ref{sec:Conclude}  concludes.  Unless specified otherwise, we use natural units, setting $c=k_b=\hbar=1$ and $\epsilon_0=\frac{1}{4\pi}$.
\\

\section{The Blob Binary Merger Rate}
\label{sec:Rates}

Dense blobs form a binary system if their mutual attraction can overcome the Hubble expansion pulling them apart.  Tidal forces from the next nearest blob prevent the pair from colliding head on and lead to the formation of a binary.  These binaries emit gravitational radiation, which leads them to in-spiral and collide, generating large amounts of energy and, perhaps, an observable signal. The binary merger rate at arbitrary times can be found by modifying an approach used for PBH binaries and outlined in Refs.\cite{Nakamura:1997sm,Sasaki:2016jop}.   

The subsection below describes a general approach to estimating the blob merger rate and provides a specific estimate for the merger rate of cold point-like blobs that primarily interact  through gravity.  Accounting for suppression effects described in later subsections, the merger rate in the Milky Way is about $\sim10^{-3} (M/10^{15}g)^{-32/37}$ pc$^{-3}$ yr$^{-1}$ for blobs of mass $M$ when blobs compose a fraction $f=1$ of dark matter. 

Subsection \ref{subsec:multi}  details how gravitational interactions between binaries and other blobs and matter perturbations reduce the merger rate.  When $f\sim1$, blobs form multibody clusters, which interact with and disrupt binaries.  When $f\ll1$, blobs do not form multibody clusters, but scattering with individual blobs can still disrupt binaries and tidal forces from other blobs and matter perturbations can deform them. These effects reduce the merger rate by a factor of $\sim3\times 10^{-4}$  for blobs merging in the present when $f\sim 1$.  This suppression weakens by at least two orders of magnitude at earlier times and when $f\ll1$. 

Subsection \ref{subsec:finite} describes how the merger rate estimates change when the blobs cannot be treated as point-like.  Binaries merge much earlier than predicted when the size of the blobs is similar to that of their orbits.  A blob's finite size can be ignored as long as its radius $r_b<10^{9}(M/10^{15}\text{g})^{-8/37} r_s$, where $r_s=2GM$ is the Schwarzschild radius of a black hole of mass $M$.  $G$ is the gravitational constant.   
Subsection \ref{subsec:friction} will describe how to account for friction between the blob and the thermal bath,  and Subsection \ref{subsec:charge} will outline how to set up the estimate for the merger rate for charged blobs.
\\

\subsection{Point-like Gravitationally Interacting Blobs }
\label{sec:Ratesmain}

We assume the blobs to have formed and decoupled from the SM at very early times.  Thus, deep in radiation domination, blobs were tightly packed together and moving with the Hubble flow.  This is when the binaries merging in the present and observable past originally formed. We assume blobs are effectively stationary relative to the thermal background, but will touch on the case when blobs are thermally coupled toward the end of this section.    The initial comoving separations between blobs can be described by a Poisson distribution. In this work, comoving distances refer to the separation blobs would have in the present, when the scale factor is normalized to 1, if they never decouple from the Hubble flow.  Pairs of blobs which happen to be close together may decouple from the Hubble flow and begin free falling toward each other.  The tidal force of the next nearest blob perturbs the  in-falling pair and prevents a head on collision, instead producing an eccentric binary.  The new binary loses energy through gravitational radiation until the blobs collide. The orbital parameters for any individual binary, its semi-major axis, $a$, and its eccentricity, $e$, depend on the mass $M$, the initial comoving distance between the blobs which form the binary, $x$, and the comoving distance to the next nearest blob outside of the binary, $y$.    Together, $a$, $e$, and $M$ dictate the binary's in-spiral time, $t$.  The spatial distribution of blobs determines the initial distribution of orbital parameters, which sets the distribution of blob merger times.
The total blob merger rate at a given time, $t$, is set by the the probability that a binary forms with the correct parameters to merge after $t$ has passed.\footnote{Binaries that merge at observable times decouple at such high redshifts that we can treat $t$ as both the in-spiral time and the time in the Universe at which the merger takes place.}  The probability that two blobs are separated by $x$, and that the next nearest blob is a distance  $y$ away is
\begin{equation}
\label{eq:poisson}
    dP = (4\pi n)^2 x^2 y^2 dxdy,
\end{equation}
where $n=\frac{\rho_{dm}f}{M}$ is their average number density today and $\rho_{dm}$ is the dark matter density today.  We require $x<y$ and $y<y_{\text{max}} \equiv (\frac{4\pi}{3}n)^{-1/3}$, the average separation between blobs.

Blobs decouple from the Hubble flow and form a stable binary when they can free fall into each other in less than the Hubble time. A binary's decoupling redshift, $z_d$, determines $a$:
\begin{equation}
\label{eq:gravsemi}
   a=\frac{x}{2(1+z_d)}.  
\end{equation}
During radiation domination, equating blob in-fall time with the Hubble time gives
\begin{equation}
    \label{zd}
     (1+z_{d}) = \frac{3M(1+z_{eq})}{8\pi \rho_{dm} x^3},
\end{equation}
where $z_{eq}$ is the redshift at matter radiation equality\footnote{One can get more precise estimates by numerically determining the decoupling redshift as in Refs.\cite{Ali-Haimoud:2017rtz,Raidal:2018bbj}, but this should only alter the result by an order (1) factor at the cost of significantly complicating the calculation.}. For the most eccentric blob binaries merging at times $t \lesssim 10^{84} t_0(M$/1 g)$^{-5/3}f^7$, $z_d \sim 3\times 10^9$ ($M$/1 g)$^{-10/37} f (t/t_0)^{-3/37}$, with $t_0$ being the age of the Universe today.  The blobs should be fully formed by $z_d$.

The eccentricity, $e$, is set by the tidal force of the next nearest blob on the binary.  This perturbs the in-falling blobs, which prevents a head on collision, producing a highly eccentric binary instead. The semi-minor axis of the binary, $b$, can be found, as in Ref.\cite{Nakamura:1997sm}, by taking  (blob free fall time)$^2\times$(tidal force from nearest blob).
\begin{equation}
    b=\frac{x^3}{GM(1+z_d)^3}\frac{GMx(1+z_d)^2}{y^3}= \left(\frac{x}{y}\right)^3a
\end{equation}
This determines the eccentricity.
\begin{equation}
\label{eq:gravecc}
    e=\sqrt{1-\left(\frac{a}{b}\right)^2} = \sqrt{1-\left(\frac{x}{y}\right)^6}
\end{equation}

The time it takes a black hole binary to merge due to gravitational radiation is applicable to point-like blobs \cite{PhysRev.136.B1224}:  
\begin{equation}
\label{eq:mergetimegrav}
    t = \frac{3}{170}\frac{a^4}{G^3M^3}\left(1-e^2\right)^{7/2},
\end{equation}
where $t$ is shown in the limit that $e\sim 1$, as the most eccentric binaries dominate the population merging at observable times.  One can find the fraction of blobs with starting eccentricity $e$  that merge at $t$ by using  \eqref{eq:gravsemi},  \eqref{eq:gravecc}, and \eqref{eq:mergetimegrav} to solve for $e$ and $t$ in terms of $x$ and $y$ and then substituting these expressions into \eqref{eq:poisson}: 
\begin{equation}
     dP = Cf^2\left(\frac{M}{ t}\right)^{5/8}\frac{e}{(1-e^2)^{45/16}}dt de,
\end{equation}
where $C=\left(\frac{27G^3}{170}\right)^{3/8}\frac{(\pi\rho_{dm}(1+z_{eq})^3)^{1/2}}{64}$. The fractional merger rate per blob can be found by integrating this over all $e$ capable of producing a binary that merges at $t$.
\begin{equation}
\label{eq:exmergerate}
    \Gamma_{blob} = \frac{8}{29}C
    f^2\left(\frac{M}{ t}\right)^{5/8}\left[\frac{1}{(1-e^2_{\text{upper}})^{29/16}}-1\right]\times S,
\end{equation}
where we've included $S$, a suppression term that accounts for disruptions to the binary by matter perturbations and external blobs, which we will  determine in Section \ref{subsec:multi} and should be no smaller than $3\times 10^{-4}$ for most $M$ and $f$. The parameter $e_{\text{upper}}$ is the largest possible eccentricity for a binary that merges in $t$, which we now discuss.  

The merger rate is dominated by the most eccentric binaries, making it sensitive to $e_{\text{upper}}$.  This maximum $e_{\text{upper}}$ is set by either the eccentricity corresponding to $y_{\text{max}}$, or the eccentricity of the binary that decouples at $z_{eq}$ and merges at $t$, whichever is smaller.  The next nearest blob to a binary, the one that perturbs it and sets its eccentricity, can be no more than $y_{\text{max}}$ away, and  $e$ grows with $y$, so it is limited by $y_{\max}$.  The eccentricity is also limited by the latest possible decoupling time, $z_{eq}$.  During matter domination, the Hubble time grows as quickly as the blob in-fall time and new binaries cannot form by the mechanism described above.\footnote{Binaries can still form during matter domination through a different mechanism,  gravitational wave bremsstrahlung during close encounters between blobs \cite{Kritos:2021nsf}} Those that decouple later---closer to matter radiation equality---get pulled further apart by the Hubble flow and must have higher eccentricities in order to merge at observable times.  
\begin{equation}
\label{eq:euppergrav}
    e_{\text{upper}}=
    \begin{cases}
    \sqrt{1-\left(\frac{t}{T_y}\right)^{6/37}} &  t<t_c\\
    \sqrt{1-\left(\frac{t }{T_z}\right)^{2/7}} & t\geq t_c\\
    \end{cases}
\end{equation}
where 
\begin{equation}
\begin{aligned}
    T_z =\frac{ks^{4/3}}{(1+z_{eq})^4}&\text{,  }\quad \quad \quad \quad T_y = k \left(\frac{ 2 y_{\text{max}}^{4}}{s(1+z_{eq})}\right)^4 \text{, }\\ \text{and } t_c = & \left(\frac{ks^{4/3} }{(1+z_{eq})^4}\right)^{37/16}\frac{1}{T_y^{21/16}}.\end{aligned}
\end{equation}  
$k = \frac{3}{170G^3M^3}$ and $s = \frac{3M}{8\pi\rho_{dm}}$.  The first term in \eqref{eq:euppergrav} is set by $y_{\text{max}}$, the second by $z_{eq}$.  

For a sense of scale, consider binaries composed of $10^{15}$ g blobs that merge today.  As the most eccentric binaries dominate the merger rate, a realistic example binary would have $e=e_{\text{upper}}=1-4\times 10^{-10}$.  It would decouple when $z\sim 2\times 10^7$, and have an initial semi-major axis of  $\sim4$ km.  The next nearest blob would be $\sim250$ km away when the binary decouples.  We can estimate the fractional merger rate per blob today using \eqref{eq:exmergerate} by taking $f=1$ and noting that $t_c$ is greater than the age of the Universe for most blobs of interest.  
\begin{equation}
    \Gamma_{blob}=10^{-17}\left(\frac{M}{ 10^{15}g}\right)^{5/37}\text{ yr}^{-1}
\end{equation}
Multiplying this by the local blob number density, assuming the local dark matter density is $\rho_{dm}^{loc} = 0.3$GeV cm$^3$, gives the local merger rate:
\begin{equation}
\label{eq:gravrateestimate}
   R_{loc}= 10^{-3} 
   \left(\frac{M}{10^{15}g}\right)^{-32/37}  \text{  yr}^{-1} \text{ pc}^{-3} 
\end{equation}
The local merger rate's dependence on $f$ is somewhat complicated because the suppression term $S$ has a complicated dependence on $f$.  For rough estimates one can take $R_{loc}\propto f^{2/5}$ when $f\sim 1$ and  $R_{loc}\propto f^{2}$ when $f\ll 1$. The merger rate in the entire Milky Way can be estimated by integrating (\ref{eq:gravrateestimate}) over the volume of the Milky Way.  The resulting merger rate will depend on one's choice of dark matter profile.  We estimate it using an NFW halo profile, where the dark matter density is described by $\rho(r) = \frac{\rho_0}{\frac{r}{h}\left(1+\frac{r}{h}\right)^2}$, and where we take $\rho_0$ = 0.4 GeV cm$^{-3}$, $h=12.5$kpc \cite{Sofue_2012}, and $r_{MW}\sim 300 $ kpc to be the radius of the Milky Way halo and the maximum distance included in the integral \cite{Eisenhauer:2003he,Deason_2020}.  The merger rate in the Milky Way halo is then $R_{MW}= 9 \times 10^{4}\left(\frac{M}{10^{15}g}\right)^{-32/37}  \text{  s}^{-1}$.

An important assumption is that the blobs are cold and have negligibly small peculiar velocities at $z_d$.  For the most eccentric blob binaries merging at time $t <t_c\sim 10^{84} t_0(M$/1 g)$^{-5/3}f^7$, $z_d \sim 3\times 10^9$ ($M$/1 g)$^{-10/37} f (t/t_0)^{-3/37}$.  If the blobs have too much kinetic energy at $z_d$ then the binaries they form will have more angular momentum and a smaller $e$ than predicted.  This causes most binaries that would have merged today to merge much later, suppressing the observable merger rate.  This is avoided if the blobs forming the most eccentric binaries contribute less angular momentum to them than tidal forces from the next nearest blob or, if the individual blobs merging at time $t<t_c$ have an average kinetic energy $< 2 \times 10^{-14}$ GeV $(M$ /1 g)$^{65/37} f^{60/37} (t/t_0)^{2/37} $.  

This kinetic energy limit impacts how blobs can be coupled to the SM background.  Blobs with $M>3\times 10^6 f^{-1/2}$ g may be thermally coupled to the SM background at $z_d$ without it interfering with their present day merger rate.  Blobs with $2\times 10^{-5}  f^{-1/6} \text{ g} <M <3\times 10^6f^{-1/2}$ g can form binaries which merge today as long as they decouple from the thermal bath early enough to redshift their excess kinetic energy away by $z_d$.  These blobs must decouple from the thermal bath by $z\simeq2\times 10^{22}(M/\text{g})^{-75/37}f^{-1/2}$ to merge at the predicted rate today.  Blobs with $M<2\times 10^{-5}f^{-1/6}$ g  will have excessive velocities at $z_d$ if they are ever thermally coupled to the SM background.  These blobs will only merge at the predicted rate if they form cold and never thermalize with the background. 


\subsection{Suppressions from Clustering and Three Body Interactions}
\label{subsec:multi}

Most binaries merging at observable times are highly eccentric and have small angular momenta.  This makes them sensitive to interactions with other blobs and matter perturbations, which tend to drive up their angular momenta and reduce their eccentricities \cite{Raidal:2018bbj}.  Given that $t\propto (1-e^2)^{7/2}$, a small reduction in  $e$ when $e\sim1$ can increase the in-spiral time by many orders of magnitude.  Interactions with other blobs and matter perturbations can therefore reduce the number of binaries merging at observable times.  Which interactions most strongly affect the merger rate primarily depend on $f$.  When $f\sim1$, blobs form dense, self-gravitating multibody clusters during matter domination.  Most binaries that fall into these clusters scatter repeatedly with other blobs, which drives $e$ down and prevents them from merging at observable times.   When $f\ll 1$, blobs do not form dense clusters, instead binaries are most affected by tidal forces from other blobs and matter perturbations, and by scattering events with individual blobs.  All of these effects suppress the observed blob merger rate.  We estimate the size of this suppression in the $f\sim 1$ case, $S_c$, and the $f\ll1$ case, $S_3$, below.  The overall suppression, $S$, corresponds to whichever of these effects dominates for a given $f$, $M$, and $z$. 

In the limit $f\sim1$, Poisson fluctuations create blob number density perturbations and overdensities.  During matter domination, these Poisson overdensities grow and eventually collapse, forming dense self-gravitating clusters.   Simulations of similar PBH clusters and binaries suggest that most blob binaries scatter repeatedly with and are disrupted by other blobs when they fall into a cluster \cite{Raidal:2018bbj,Jedamzik:2020ypm}.  However, not all clusters are equally destructive to binaries.  As the number of blobs in a Poisson cluster, $N$, increases, the blob number density decreases, and the cluster tends to become gravitationally bound at later times \cite{Jedamzik:2020ypm}.   Binaries will not be damaged by a cluster which has not yet collapsed.  To account for this, we assume binaries only remain in tact and able to merge at their predicted time if they either never fall into a cluster or fall into one which has not yet become self-gravitating at the time of the merger.\footnote{This is a conservative assumption, not all binaries in clusters are destroyed \cite{Jedamzik:2020ypm}, and new binaries capable of merging within the lifetime of the Universe may form within clusters \cite{Kritos:2021nsf}.}  

We estimate $S_c$ as the probability that a binary is either not in a cluster or in a cluster too large to have begun self-gravitating.  Given a collection of $N_{max}$ Poisson distributed objects, the probability that one is in an $N$ object cluster is \cite{1983MNRAS.205..207E}  \begin{equation}
\label{eq:probability}
    p_{N,N_{max}} = \frac{N^{-1/2}e^{-N/N^*}}{\sum_{N=2}^{N_{max}}N^{-1/2}e^{-N/N^*}},
\end{equation}
where $N^*$ is the characteristic scale of clusters at $z$ as determined analytically by Ref.\cite{1983MNRAS.205..207E} and confirmed by simulations in Ref.\cite{Inman:2019wvr}.
\begin{equation}
    N^* =\left(\text{ln}(1+\delta_*)-\frac{\delta_*}{(1+\delta_*)}\right)^{-1},
\end{equation}
where $\delta_*$ is the size of an initial density perturbation which collapses at $z$.
\begin{equation}
    \delta_* = \frac{1.68}{D f}
\end{equation}
D is a growth function which describes how an initial density perturbation evolves during matter domination.  An initial perturbation, $\delta$, grows approximately linearly with $z$, and collapses once $\delta = 1.68$. Ref.\cite{Inman:2019wvr} found  
\begin{equation}
    D\equiv \left(1+\frac{7}{5}\frac{(1+z_{eq})}{(1+z)}\right)^{0.9}.
\end{equation}
Using \eqref{eq:probability} one can find, $p_{2,\infty}$, the probability that a blob is part of a ``cluster'' of 2---a binary---and $\sum_{N'=N_c}^{\infty} p_{2,N'}p_{N',\infty}$, the probability that a blob is in a binary inside of a cluster made of at least $N_c$ blobs.  If we take $N_c$ to be the smallest cluster which has not collapsed by $z$, then this sum describes the fraction of binaries in clusters that will not collapse by $z$. Poisson density perturbations of $N$ blobs grow as $D\frac{1}{\sqrt{N}}$ during matter domination, and collapse once $1.68=D\frac{1}{\sqrt{N}}$.  This makes $N_c = \left(\frac{D}{1.68}\right)^2$.  Together, the two above probabilities give $S_c$, the total likelihood that a binary will not be in a cluster which collapses around it before the merger happens. 
\begin{equation}\label{eq:Sc}
    S_c = p_{2,\infty}+\sum_{N'=N_c}^{\infty} p_{2,N'}p_{N',\infty}.
\end{equation}
$S_c\sim 3\times 10^{-4}f^{-1}$ when $f\sim1$ for mergers happening today.

When $f\ll1$,  binaries are primarily disrupted by two different interactions; tidal forces from other blobs and matter density perturbations at the binary's formation, and scattering with a third blob during matter domination. Tidal forces from other blobs and matter density perturbations can decrease the a binary's eccentricity while it's forming, which ultimately reduces the binary merger rate in the present.  This effect is only significant when $f\ll1$. While Poisson fluctuations will not lead to the growth of blob clusters when $f\ll1$, binaries themselves can behave as small overdensities which grow and collapse during matter domination.  A binary will be disrupted if a third blob is close enough to be pulled into this overdensity.   The merger rate suppression from these two effects, $S_3$, was estimated in Ref.\cite{Raidal:2018bbj} for a monochromatic mass distribution of PBHs and is applicable to blobs. 
\begin{equation}
    S_3 \simeq 1.4\left(\frac{1}{N_d}+\frac{\sigma_M^2}{f^2}\right)^{-21/74}e^{-N_d}
\end{equation}
The sum term above accounts for the suppression from tidal interactions with external blobs beyond the next nearest one which sets the initial eccentricity, $(1/N_d)$, and matter density perturbations, ($\sigma_M^2/f^2$), during blob formation. The exponential term excludes binaries with a third blob close enough to disrupt it.  $N_d$  is the average number of blobs close enough to the binary to disrupt it.  $\sigma_M$ is the root mean square scale of matter perturbations at the mass scale of the blob  \cite{Eroshenko:2016hmn}:
\begin{equation}
    \sigma_M = 8.2\times10^{9.25(n_s-1)-3}\left(\frac{M}{\text{g}}\right)^{\frac{1-n_s}{6}}\times \left[1-0.06\text{ Log}_{10}\left(\frac{M}{2\times10^{33}\text{g}}\right)\right]^{3/2}
\end{equation}
where $n_s = 0.965$ is the scalar spectral index \cite{Zyla:2020zbs}. 

  We estimate $N_d$, as in Ref.\cite{Raidal:2018bbj}, as the average number of blobs in the overdense region that collapses around the binary.  We assume any blob in this region can disrupt the binary.   The binary creates an initial overdensity $\delta_0 = \frac{2M-V\rho_{dm}f}{V\rho_{dm}}$ in a comoving volume $V$ around it.  During matter domination, this overdensity grows as $\delta\simeq \delta_0D$  and collapses once $\delta = 1.68$ \cite{Inman:2019wvr}. 
Therefore, any initial overdensity $\delta_0>\delta_I\equiv \frac{1.68}{D}\simeq 3.5\times 10^{-4}(1+z)$ will collapse by $z$.  Noting that $N_d=V\rho_{dm}f$, one can rearrange the expressions for $\delta_0$ and $\delta_I$ to find $N_d$.
\begin{equation}
    N_d = \frac{2f}{\delta_I(z)+f}
\end{equation}
For mergers happening today $S_3\sim 0.08$ for $M=10^{15}$g and $f=10^{-4}$. Overdensities do not grow significantly outside of matter domination.

Whether 3 body scattering and tidal perturbations or clustering effects dominate the merger suppression depends on $z$, $M$, and $f$. Ref.\cite{Inman:2019wvr} estimates that the majority of PBHs are part of multibody clusters when $f\gtrsim 2\times 10^{-4}(1+z)$, while most are isolated or part of binaries for smaller $f$. Modeling will be needed to estimate precisely how much clustering suppresses the binary merger rate when $1>f>2\times 10^{-4}(1+z)$. One can reasonably expect it to be no less than the smaller of $S_c$ and $S_3$.  We  estimate 
\begin{equation}
    S=\begin{cases}
    S_c& f=1\\
    S_3& f<2\times10^{-4}(1+z)\\
    \text{Min}[S_c,S_3]&1> f>2\times10^{-4}(1+z).\\
    \end{cases}
\end{equation}
\\
\subsection{Finite Size Effects}
\label{subsec:finite}
Section \ref{sec:Ratesmain} estimates the merger rate for point-like blobs. Blobs of radius $r_b$ collide when they come within $2r_b$ of each other, reducing the binary's in-spiral time. This leads mergers to happen more quickly than predicted in Section  \ref{sec:Ratesmain} and reduces the number that occur at observable times.

One can account for this in the merger rate estimate by only counting binaries whose in-spiral time is not significantly changed by the finite size of the blobs. This is done by adjusting  $e_{\text{upper}}$ to
\begin{equation}
\label{eq:eupper}
     e_{\text{upper}}=
    \text{Min}\left[\
    \sqrt{1-\left(\frac{t_r}{t}\right)^2}\text{, }\sqrt{1-\left(\frac{t}{Ty}\right)^{6/37}}\text{, }
    \sqrt{1-\left(\frac{t }{T_z}\right)^{2/7}} \right]
\end{equation}
with 
\begin{equation}
    t_r=\frac{24}{85G^3}\frac{r_b^4}{M^3} .
\end{equation}
The first term in \eqref{eq:eupper} excludes blobs that collide too early, while the second two come from the expression for $e_{\text{upper}}$ for point-like blobs. $t_r$ corresponds to the merger time of a binary with starting $a=2r_b$ and $e=0 $. When $t_r>t$, all blobs that would otherwise merge by $t$ collide early.

This analysis holds for mergers happening today if $t_r<t_0$, where $t_0$ is the age of the Universe.  This corresponds to \begin{equation}
    r_b<r_{\text{max}}\equiv\left(\frac{85 G^3M^3t_0}{24}\right)^{1/4} \simeq 10^{10} \left(\frac{M}{10^{15}\text{g}}\right)^{-1/4} r_s.
\end{equation}
$r_{\text{max}}$ should be larger than the blob's Compton radius $\left(\frac{2\pi}{M}\right)$.  This is the case for  $M>10^{-13}$ g or $6\times 10^{10}$ GeV.  
The blob must be a black hole to avoid finite size effects strongly suppressing the  merger rate when $r_{\text{max}}<r_s$.  This is the case for $M>4\times 10^{55}$ g or $2\times 10^{22}$M$_{\odot}$, well beyond the reasonable upper limit on the dark matter mass.  

Finite size effects do not alter the merger rate as long as the first term in \eqref{eq:eupper} is larger than one of the latter.  For most blobs merging today, when $f=1$, finite size effects can be ignored if
\begin{equation}
\label{eq:denseenough}
    r_b<r_{\text{fin}}\equiv\left(\left(\frac{t_0}{T_y^{3/40}}\right)^{40/37}\frac{85G^3M^3}{24}\right)^{1/4}\simeq10^{9}\left(\frac{M}{10^{15}\text{g}}\right)^{-8/37}r_s.
\end{equation}  $r_{\text{fin}}$ is greater than the blob Compton radius and  Schwarzschild radius for blobs with $1.5\times10^{12}\text{ g}< M  < 4\times 10^{56} $ g.  The upper limit is again well beyond the reasonable mass range for dark matter, and thus a large range of densities are allowed without finite-size effects disrupting the merger rate.
\\

\subsection{Friction Effects}
\label{subsec:friction}
Long-range, non-gravitational interactions between the blob and the SM create friction with the thermal bath.  This can alter the behavior of blob binaries by delaying their decoupling from the Hubble flow, removing energy from the orbit, and causing the blobs to thermalize with the SM, which may leave them too energetic to form binaries as predicted.  

Friction slows blob movement, reducing their ability to fall toward each other, and delaying their decoupling from the Hubble flow.  The frictional force will take the general form
\begin{equation}
    \frac{dE}{dx} = -F(z, M, \epsilon,v).
\end{equation}
$F(z,M,\epsilon,v)$ contains the model specific relation between the blob's velocity, $v$ and its friction with the SM. It may depend on $z$, $M$, the blob's coupling to the SM, $\epsilon$, and $v$.  Friction delays binary decoupling if it more strongly influences the blob's peculiar velocity than Hubble expansion at  $z_d$, i.e. if     $F(z_d,M,\epsilon,v)>H(z_d)M v$.  In this case, blobs decouple at $z_f$, when their terminal in-fall velocity, $v_{t}$, dictated by
\begin{equation}
    F(z_f,M,\epsilon,v_t)=\frac{GM^2}{x^2(1+z_f)^2}
\end{equation} exceeds the Hubble expansion velocity moving them apart $\frac{H(z_f) x}{1+z_f}$. One can find $a$  and $e$ for these binaries by replacing $z_d$ with $z_f$ in  \eqref{eq:gravsemi} and \eqref{eq:gravecc}, respectively.

The binary in-spiral time may be reduced if friction removes energy from the binary more efficiently than gravitational radiation. Estimating this requires a specific model for the friction with the thermal bath. As indicated in Ref.\cite{Diamond:2021scl}, friction between the blobs and the thermal bath strongly suppresses the merger rate.  One can reasonably approximate the merger rate by only counting binaries not strongly influenced by friction, those for which $F(z_d,M,\epsilon,v)<H(z_d)M v$.

One can show that for blob sizes allowed in Section \ref{subsec:finite}, even a geometric cross section between the blobs and SM radiation will not alter the mergers rate, which long-range interactions can.   While friction with the thermal bath will not prevent blobs with short range interactions with the SM from decoupling at their usual time, it can still thermalize them.  As mentioned in Section \ref{sec:Ratesmain} thermalized blobs may have enough kinetic energy to alter the dynamics of highly eccentric binaries and suppress the merger rate.
\\

\subsection{Charged Blobs}
\label{subsec:charge}

One can imagine models of dense blobs that self-interact through a U(1) charge---either a SM EM charge (though charged dark matter is strongly constrained \cite{Stebbins:2019xjr,Diamond:2021scl,Fedderke:2019ajk}) or a dark equivalent.  While less general than models that solely interact through gravity, these are worth considering  because they merge more quickly than their uncharged counterparts.  We note that, charged interactions likely change the way external blobs  cluster and disrupt binaries from what was described in Section $\ref{subsec:multi}$.  N-body simulations of binaries interacting with other charged blobs in an expanding background, similar to the work done for black hole binaries \cite{Inman:2019wvr,Raidal:2018bbj,Jedamzik:2020ypm}, are needed to understand how interactions with other blobs modify the merger rate. We leave such simulations to future work and simply outline how one would estimate the merger rate for charged blobs here. 

The charged merger rate can be found by adjusting some details of the calculation laid out in Section \ref{sec:Ratesmain}.  Charged binaries will decouple earlier, have different $e$, and merge faster than uncharged ones.  It is convenient to describe the U(1) attraction between blobs in comparison to the gravitational attraction.  
\begin{equation}
    q\equiv \frac{Q}{Q_{ext}} = \frac{Q m_p}{M}
\end{equation}
is the blob charge in terms of the charge of an extremal black hole of mass $M$. We take  $m_p=1.2\times10^{19}$g as the Planck mass, and $Q$ as the charge of the blob.  If $q=1$, the U(1) and  gravitational forces between blobs will be equal in magnitude.  We will work in the limit $q\gg1$.

Charged blobs only pair with blobs of opposite sign. Assuming an equal distribution of charges, we introduce a factor of $\frac{1}{2}$ into Eq.\eqref{eq:poisson} to account for this.  The binary's decoupling redshift increases due to the stronger attraction between blobs, becoming
\begin{equation}
 \label{zdu}
     (1+z_{q}) = \frac{3M(1+z_{eq})}{16\pi \rho_{dm} x^3}q^2.
\end{equation}

The eccentricity of the binary is set by dipole torques from the next nearest blob.  When charged interactions dominate, the next nearest blob gives the binary a semi-minor axis of (dipole torque)$\times$(free fall time)$^2$,
\begin{equation}
    b=\frac{GM(1+z_q)^2q^2}{y^2}\frac{x^3}{GM(1+z_q)^3q^2}= \left(\frac{x}{y}\right)^2a,
\end{equation}
and an eccentricity
\begin{equation}
\label{eq:U1ecc}
    e= \sqrt{1-\frac{x^4}{y^4}}.
\end{equation}
The coalescence time for charged binaries also changes (decreases) because they emit dipole radiation instead of quadrupole radiation, as in the gravitational case. The merger time for $q=1$ is derived in Ref.\cite{Diamond:2021scl} for extremal magnetic black holes and is modified here to account for $q\gg1$.
\begin{equation}
\label{eq:U1time}
    t=\frac{0.034a^3}{4G^2M^2}\frac{(1-e^2)^3}{q^4}
\end{equation}

These modified expressions for $z_q$, $e$, and $t$ can be used to estimate the merger rate for charged blobs in the same manner as was done for uncharged ones in Section \ref{sec:Ratesmain}.  A full estimate will require a more detailed investigation of how charged blobs cluster and perturb binaries and the suppression, $S$, that results from it, though we leave this to future work. 
\\

\section{An Observable Signal} 
\label{sec:Signal}
Even accounting for suppressions from multibody interactions, finite size effects, and friction, there is plenty of parameter space where blob binaries merge at a high rate today.  For comparison, the random collision rate in the Milky Way is about
\begin{equation}
    R_{rand} = \sigma v\left(\frac{\rho_{dm}^{loc}f}{M}\right)^2,
\end{equation}
where $v=10^{-3}$ is the virial velocity of blobs in the Milky Way and $\rho_{dm}^{loc}=0.3$ GeV cm$^{-3}$ is the local dark matter density.  $\sigma$ is the interaction cross section. It is the larger of the geometric cross section $\sim\pi r_b^2$ and the binary capture cross section through gravitational wave emission $\sim\frac{4.9 \pi r_s^2}{v^{18/7}}$  \cite{Kritos:2021nsf}.  The second cross section gives $R_{rand}\simeq  10^{-25} f^2$ pc$^{-3}$ yr$^{-1}$ in the Milky Way, well below the predicted rate for binary blob collisions from Eq. \eqref{eq:gravrateestimate} for most $M$.  As the random collision rate depends on the dark matter density squared, it may dominate in very dense regions such as galactic centers, though it will be small overall compared to the binary merger rate.

The observable signal produced in binary collisions will vary between blob models. If the blob has any coupling to the SM, then the high temperatures and large volumes of the collisions may produce a large flux of EM particles and neutrinos. Composite blobs, such as those described in Refs.\cite{Brandt:2016aco, Hardy:2015boa,Wise:2014jva, Gresham:2017zqi,Bai:2018dxf}, generally produce particles with energies similar to their binding energy.   Collisions that form low mass black holes will produce radiation with a spectrum described in Ref.\cite{Ukwatta:2015tza}. Even blobs totally decoupled from the SM will produce a gravitational wave signal.    

The merger rate scales linearly with the local dark matter abundance, generating excess signal from galactic halos. Neglecting attenuation effects, the flux of energetic particles produced in binary collisions in the Milky Way can be estimated by
\begin{equation}
\begin{split}
    F_{gal}&=\int_0^{r_{MW}+r_{\odot}}R\frac{\rho(r)f}{4\pi M}*2M\chi_i ds d\Omega \\
    &=2\times 10^{-3}\left(\frac{M}{g}\right)^{5/37} f^{53/37} \chi_i  S \text{ GeV cm}^{-2}\text{ s}^{-1}.
\end{split}
\end{equation}
$\rho(r) = \frac{\rho_0}{\frac{r}{h}\left(1+\frac{r}{h}\right)^2}$ is the NFW density profile for the Milky Way with $\rho_0$ = 0.4 GeV cm$^{-3}$ and $h=12.5$kpc \cite{Sofue_2012}. $r_{\odot}= 8$ kpc is the distance between the Sun and the center of the Galaxy, and $r_{MW}\sim 300 $ kpc is the radius of the Milky Way \cite{Eisenhauer:2003he,Deason_2020}.  $\chi_i$ is the fraction of the blob's mass released as particle specie $i$.   $\vec{s} = \vec{r}-\vec{r}_{\odot}$ is the distance between the Earth and the blob collision.
 
One can make a similar estimate for the extragalactic flux of particles with energy $E$.   
\begin{equation}
\label{eq:extragal}
    F_{extragal}(E) = \frac{\rho_{dm}f}{2\pi}\int_0^{z_{max}}\frac{R(z)\chi_i(E(1+z))A_i(E(1+z),z)}{H(z)}dz
\end{equation}
$z_{max}$ is the greatest redshift from which particles radiated by the merger can reach Earth with energy $E$. $A_i(E(1+z),z)$ is an attenuation function which accounts for the fraction of particles emitted at redshift $z$ with energy $E(1+z)$  that reach Earth. It depends heavily on the specific particles radiated and their energies.  We describe some model-independent signals binary mergers can produce below.  
\\
\subsection{Electromagnetic Signal}
Blob mergers that produce EM particles lead to a variety of observable signals.  Most depend on the energies of the particles produced.  These may include  synchrotron radiation from the Galactic center, contributions to the 511 keV gamma-ray excess \cite{Weidenspointner:2007rs}, and cosmic rays (perhaps even contributions to the positron cosmic ray excess \cite{PhysRevLett.110.141102}).  

There are a few instances in which EM radiation is quickly reprocessed and the observable signal does not depend strongly on the energy of the original particles involved.  The first is the CMB anisotropy angular power spectrum.  EM radiation with $E>10$ keV released around recombination drives up the free electron fraction, distorting the anisotropy angular power spectrum \cite{Poulin:2016anj}.  The bound on radiation emission is strictest at $t\sim10^{13}$ s or $z\sim 1200$, where it is limited to $\sim 2\times 10^{-20}$ GeV cm$^{-3}$ s$^{-1}$.  Using Eq.\eqref{eq:exmergerate} we find blobs that satisfy
\begin{equation}
   3\times 10^{-3}<\left(\frac{M}{g}\right)^{5/37} f^{53/37}\chi_{EM}\times S 
\end{equation}
 are ruled out by observations of the CMB.

\begin{figure}
    \centering
    \includegraphics[scale=0.7]{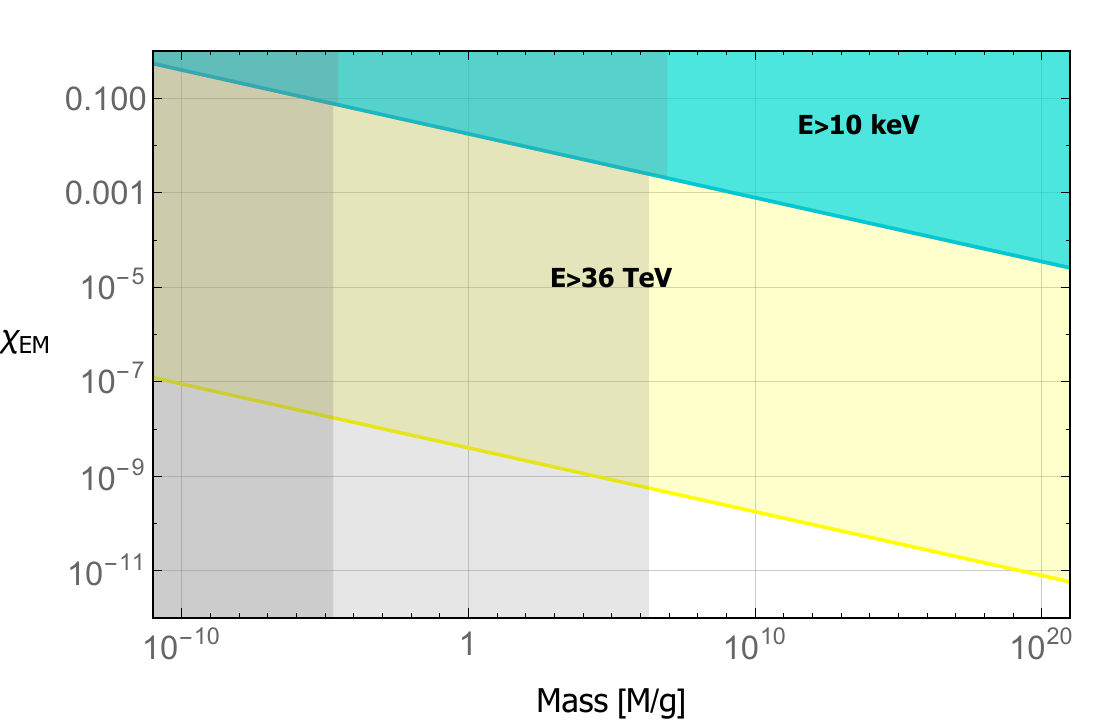}
    \caption{Limits on $\chi_{EM} $ for blobs that satisfy \eqref{eq:denseenough} when $f=1$.  The yellow region excludes $\chi$ for EM particles with $E>36$ TeV.  These are reprocessed into gamma-rays whose flux would be greater than that observed by Fermi-Lat.  The blue region excludes $\chi$ for EM particles with $E> 10$ keV.  These increase the free electron fraction and distort the CMB anisotropy angular power spectrum if emitted at $z\sim 1200$. Blobs with $M>10^{21}$g and $f=1$ are ruled out by lensing searches \cite{Niikura:2019kqi}.  The gray regions show where thermal couplings to the SM bath can give blobs too much kinetic energy at $z_d$ to produce binaries as expected.  This suppresses the merger rate and is described in Section (\ref{sec:Ratesmain}). Blobs in the dark gray region only produce a signal if they  never thermally couple with the SM. Those in the light gray region produce a signal as long as they thermally decouple from the SM bath early enough to redshift away their excess kinetic energy.  The remaining blobs will produce the expected signal even if they remain thermally coupled to the SM bath through $z_d$.  The gray regions are slightly offset for the different signals because these come from mergers taking place at different times.  The binaries involved in these mergers have different $z_d$ and different sensitivities to the blob temperature at $z_d$.}
    \label{fig:EMlim}
\end{figure}

The second signal  that does not depend strongly on the energy of the original particles radiated is the diffuse gamma-ray background.  EM radiation with $E>E_{th}\equiv\frac{36000}{1+z}$GeV is quickly reprocessed by scattering with CMB photons.  This generates a gamma-ray spectrum of the form  \cite{Berghaus:2018zso,1989ApJ...344..551Z}
\begin{equation}
    L(E,z) =
\begin{cases}
0.767 E_{th}(z)^{-0.5}E^{-1.5} & E\leq0.04E_{th}(z)\\
0.292E_{th}(z)^{-0.2}E^{-1.8} & 0.04 E_{th}(z)<E<E_{th}(z).
\end{cases}
\end{equation}
Radiation with $E<E_{th}$ free streams to Earth  \cite{1989ApJ...344..551Z}, while that at higher energies is reprocessed. As long as $E>E_{th}$, the shape of the final spectrum is independent of the starting one.  The flux that reaches Earth can be found using Eq. \eqref{eq:extragal}.
\begin{equation}
    F(E) =  \frac{\rho_{dm}f}{2\pi}\int_0^{z_m}\frac{R(z)\chi_{EM}L(E(1+z),z)}{H(z)}dz
\end{equation}
Blobs cannot produce more gamma-rays than what has been observed by the most recent Fermi-LAT extragalactic gamma-ray survey \cite{Ackermann:2014usa}.  Limiting $F(E)$ to be less than 2$\sigma$ above the gamma-ray flux observed by Fermi, gives a bound on $\chi_{EM}$.  We take the strongest limits on $\chi_{EM}$, which come from Fermi's 580 GeV-820 GeV energy bin. This constraint is displayed in Fig.(\ref{fig:EMlim}).  
\\
\subsection{Gravitational Wave Signal}
Blobs with $M>10^{20}$g can produce gravitational wave signals relevant to LIGO, LISA, or future gravitational wave detectors.    The  gravitational wave signal from these events differs from that of black holes binaries because blobs can have $r>r_s$ and do not necessarily have event horizons.  This leads them to collide earlier and end the in-spiral phase of the merger emitting at a lower frequency than expected for a black hole binary of the same mass \cite{Giudice:2016zpa}.  The lack of event horizon causes the ring down phase immediately after the merger to differ from that of a black hole binary \cite{Giudice:2016zpa}.  When considering specific blob models,  simulations of the waveform for the binary in-spiral and merger, like those generated in Ref.\cite{Giudice:2016zpa} for ECOs, can help differentiate blob mergers from other merger events. EM signals produced in blob mergers can also help distinguish them from black hole mergers.  In general, a blob is very different from an neutron star - thus EM signals produced from their mergers should be distinct from those of neutron stars as well.  

Individual mergers taking place in the present will be observable if the amplitude of the strain, $h$, and the frequency, $\nu$, of the gravitational wave signal it produces falls within a gravitational wave detector's sensitivity range.
\begin{equation}
    h = \frac{8\pi^{2/3}}{10^{1/2}2^{1/5}d_L}G^{5/3}M^{5/3}\nu^{2/3}
\end{equation}
$d_L$ is the luminosity distance, which relates to the comoving distance, $d_c$, by $d_L = (1+z)d_c$ \cite{Sesana:2008mz}.    During in-spiral, $h$ peaks at the innermost stable circular orbit (ISCO).   For blobs of radius $r_b$, this corresponds to a frequency \cite{Giudice:2016zpa}
\begin{equation}
    \nu^{ISCO} = \frac{(GM)^{1/2}}{3^{3/2}2\pi r_b^{3/2}}.
\end{equation}

\begin{figure}
    \centering
    \includegraphics[scale=0.5]{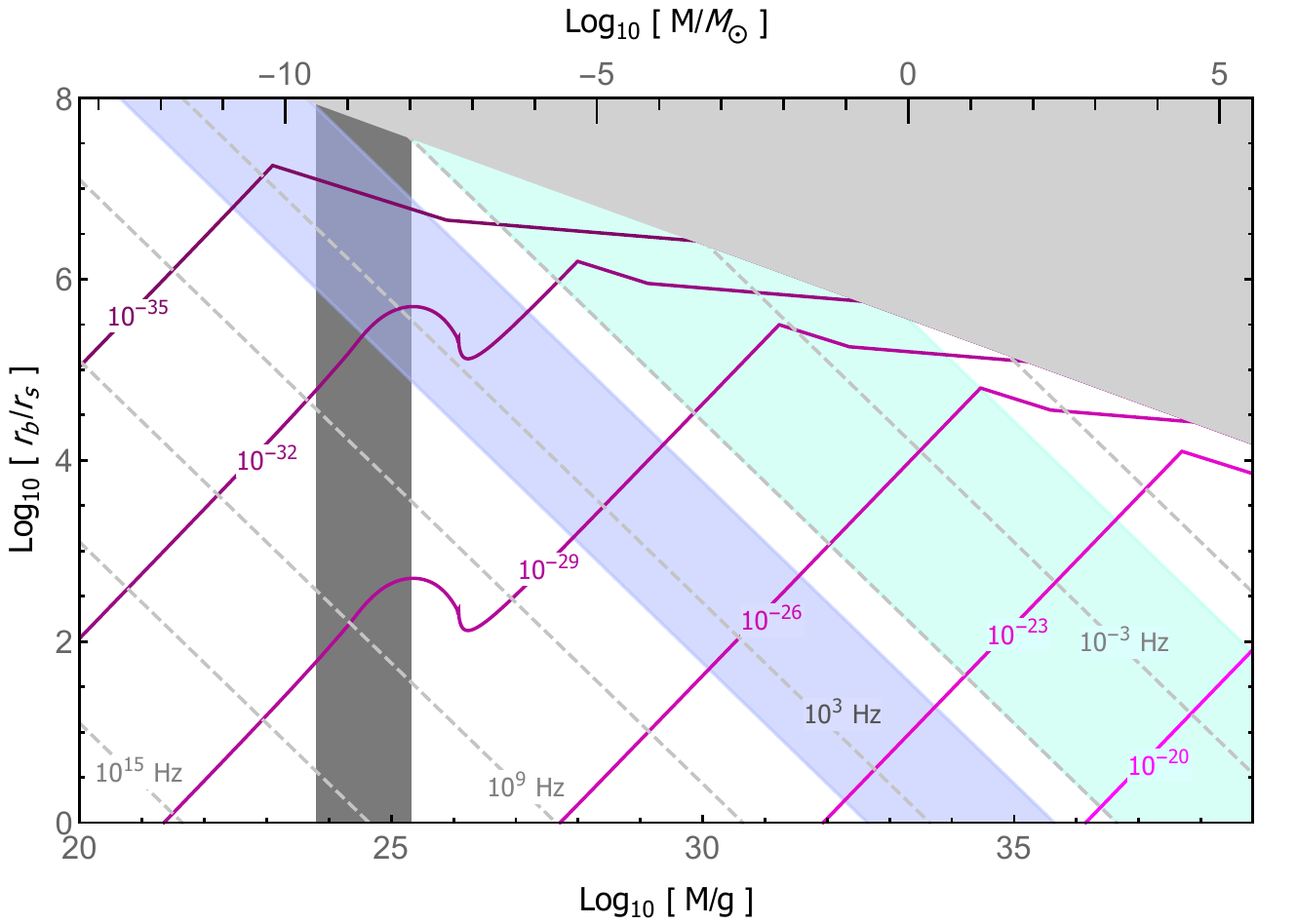}
    \caption{Contour plot showing peak strains and frequencies of gravitational waves emitted during blob in-spiral for the nearest blob merger expected in an average year when $f=10^{-2}$.  Peak strains are marked by pink lines, and peak frequencies are marked by gray dashed lines.  The upper gray region is excluded because finite size effects strongly suppress the merger rate.  The teal and purple regions highlight the frequencies LISA and LIGO are sensitive to, respectively.  The dark gray vertical band is ruled out for $f=10^{-2}$ by gravitational lensing searches \cite{Niikura:2017zjd}. The blob radius is presented in terms of the Schwarzschild radius, $r_s$.  The knee in the strain contours show where finite size effects begin to suppress the merger rate for blobs with large $r_b$.  The downward curve in the $10^{-29}$ and $10^{-32}$ strain contours shows where mergers transition from primarily occurring in the Milky Way to outside of it.  The typical distance to the nearest merger grows quickly in the transition from Galactic to extragalactic mergers, reducing the expected strain.}
    \label{fig:allstrain}
\end{figure}

For reference, we display the peak in-spiral frequency and strain of the nearest merger expected within an average year of observation time in Fig.(\ref{fig:allstrain}).  There is a large region of parameter space where blob mergers produce reasonably large strains ($10^{-29}-10^{-23}$) at frequencies above current detection sensitivities (> 10 kHz).  There are no known astrophysical sources that could produce such large gravitational wave signals in the (MHz-GHz) frequency range, making blobs a unique target for high frequency gravitational wave detectors, such as those considered in \cite{Aggarwal:2020olq}.

\begin{figure}
    \centering
    \includegraphics[scale=0.5]{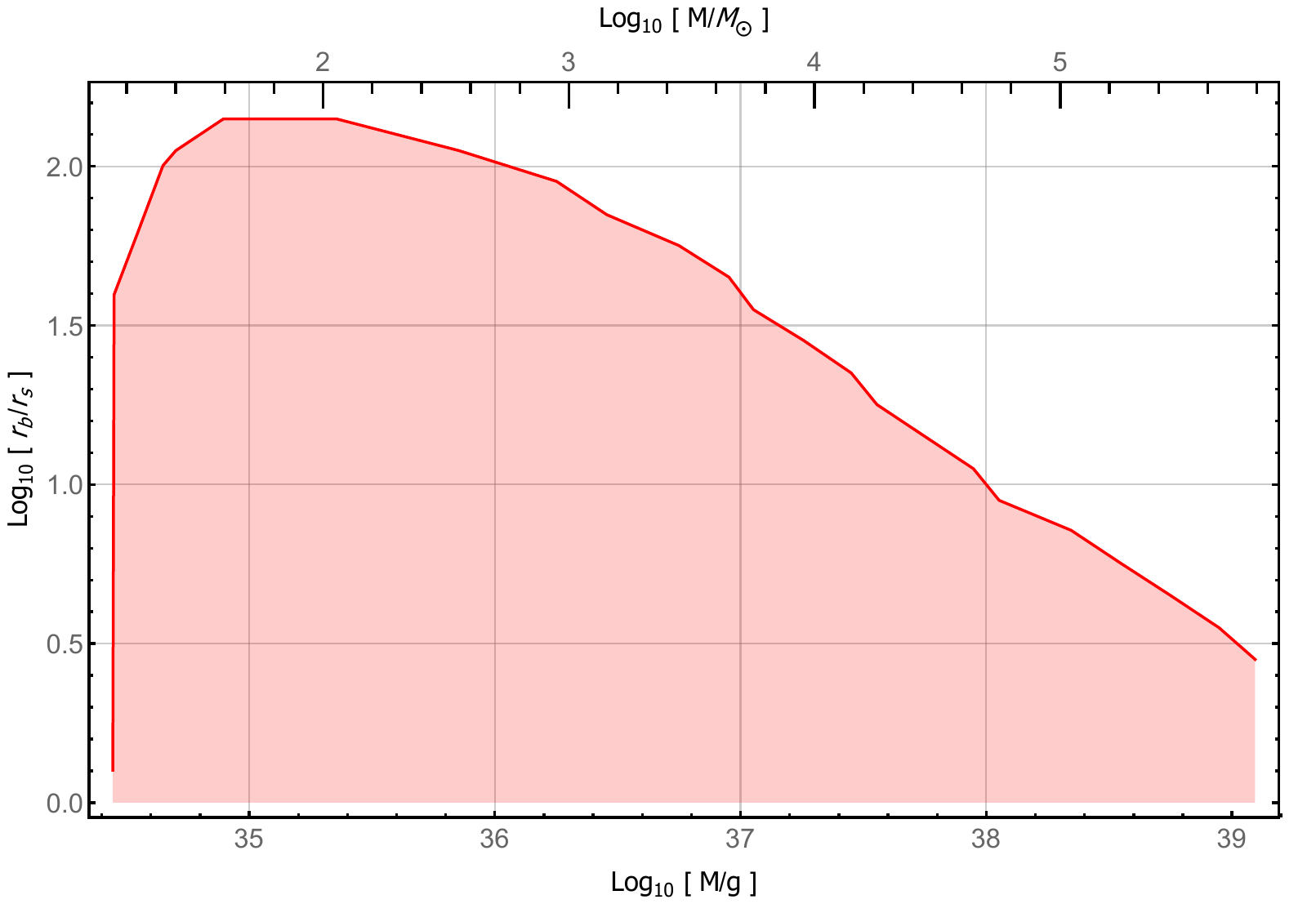}
    \caption{Parameter space where the predicted blob merger contribution to the SGB lies above LISA's expected sensitivity threshold \cite{Abbott:2016xvh} for  $f=10^{-2}$. The blob radius is presented in terms of the Schwarzschild radius, $r_s$.  The blob merger contribution to the SGB  is too small for blobs lighter than $10^{34}$g to be visible to LISA.}
    \label{fig:stocastic}
\end{figure}

Many mergers would have occurred throughout the Universe's history.  While these events would not be individually distinguishable, they would contribute to the stochastic gravitational wave background (SGB). One can compare their contribution to the SGB to detector sensitivities  by finding their combined power spectral density \cite{Christensen:2018iqi}.
\begin{equation}
    S_n(\nu)=\frac{3 H_0^2}{2\pi^2\rho_c\nu^2}\int_0^{z_{m}}\frac{R(z)\frac{dE_{GW}}{d\nu_s}(\nu_s)}{(1+z)H(z)}dz,
\end{equation}
where $\rho_c$ is the comoving critical density, and $z_{m} = \frac{\nu^{ISCO}}{\nu}-1$.   $\frac{dE_{GW}}{d\nu_s}(\nu_s)$ is the gravitational wave energy spectrum radiated by an in-spiraling binary \cite{PhysRevD.85.104024}.
\begin{equation}
    \frac{dE_{GW}}{d\nu_s}(\nu_s) = \frac{(G\pi)^{2/3}}{3}\mathcal{M}^{5/3}\nu_s^{-1/3}
\end{equation}
$\mathcal{M} $ represents the binary chirp mass, which is $M/2^{1/5}$ for an equal mass binary.  $\frac{dE_{GW}}{d\nu_s}(\nu_s)$ cuts off at $\nu^{ISCO}$, and $\nu_s=(1+z)\nu$.
We compare the predicted SGB spectra from blob mergers to LISA's expected sensitivity  \cite{Abbott:2016xvh, 2017arXiv170200786A}, and display the parameter space where blob mergers would produce an observable signal in Fig.(\ref{fig:stocastic}).  This estimate excludes contributions from binaries merging before $z_{eq}$. The decoupling time can become similar to or larger than the in-spiral time at these redshifts, altering the predicted merger rate. 
\\

\section{Conclusion}
\label{sec:Conclude}

In this work, we have demonstrated that cold dark matter heavier than $8\times 10^{11}$ GeV and dense enough to satisfy Eq.\eqref{eq:denseenough} will, model-independently, form binary pairs in the early Universe.    Many of these binaries will spin down and merge within the lifetime of the Universe, potentially producing observable signals significantly larger than those predicted for  dark matter annihilation at the same masses.  This phenomenon provides a new way to search for MACHOs and a probe of dark matter that is normally difficult to observe due to low number densities and low masses.   We determined the merger rate at arbitrary times, and have outlined how to adjust it to accommodate different blob models based on their size and non-gravitational interactions with the SM and themselves.  We have described signals that could result from these mergers, and indicated parameter space where both EM and gravitational wave signals could be observed.  We ruled out regions of blob parameter space based on observations of the CMB anisotropy angular power spectrum and gamma-rays.  We showed that merging blobs could be a unique source of high frequency gravitational waves.  More precise modeling may be needed to determine the merger rate and signals for binaries of specific blob models.  However, this work has described a novel, nearly model-independent phenomenon which sheds light on thus far unprobed regions of dark matter parameter space. 

\vspace{.2in}
\textbf{Acknowledgements:}
We would like to thank Yacine Ali-Ha\"{i}moud, Karsten Jedamzik, Joshua Kable, Ely Kovetz and Hardi Veerm\"{a}e for their helpful conversations and insights.  MD, DK, and SR are supported in part by the NSF under grant PHY-1818899.   SR is also supported by the U.S. Department of Energy, Office of Science, National Quantum Information Science Research Centers, Superconducting Quantum Materials and Systems Center (SQMS) under the contract No. DE-AC02-07CH11359. SR is also supported by the DoE under a QuantISED grant for MAGIS  and the Simons Investigator award 827042.
\bibliographystyle{JHEP}
\bibliography{ref.bib}
\end{document}